\documentclass[letterpaper]{article}
\usepackage[latin1]{inputenc}
\usepackage[T1]{fontenc}
\usepackage[english]{babel}
\usepackage{amsmath}
\usepackage{amssymb,amsfonts,textcomp}
\usepackage{color}
\usepackage{hhline}
\usepackage[normalem]{ulem}
\usepackage{hyperref}
\hypersetup{pdftex, colorlinks=true, linkcolor=blue, citecolor=blue, filecolor=blue, urlcolor=blue, pdftitle=, pdfauthor=Microsoft Office User, pdfsubject=, pdfkeywords=}

\makeatletter
\newcommand\ps@Standard{
  \renewcommand\@oddhead{}
  \renewcommand\@evenhead{\@oddhead}
  \renewcommand\@oddfoot{\textcolor{black}{\thepage{}}}
  \renewcommand\@evenfoot{\@oddfoot}
  \renewcommand\thepage{\arabic{page}}
}
\makeatother
\begin{document}
\clearpage\setcounter{page}{1}\pagestyle{Standard}

{\bf \large Basis Glass States: New Insights from the Potential Energy
Landscape \ }

\bigskip

Prabhat K. Gupta 

Dept. of Materials Science and Engineering,

The Ohio State University

Columbus, Ohio 43210, USA

\bigskip

Walter Kob

Laboratoire Charles Coulomb and CNRS

University of Montpellier

{Montpellier, F-34095, France

\bigskip

\bigskip

\bigskip

\textbf{Abstract:}\newline
Using the potential energy landscape formalism we show that, in the
temperature range in which the dynamics of a glass forming system is
thermally activated, there exists a unique set of {\textquoteleft}basis
glass states{\textquoteright} each of which is confined to a single
metabasin of the energy landscape of a glass forming system. These
basis glass states tile the entire
configuration space of the system, exhibit only secondary relaxation and
are solid-like. Any macroscopic state of the system (whether liquid or
glass) can be represented as a superposition of basis glass states and can
be described by a probability distribution over these  states. During
cooling of a liquid from a high temperature, the probability distribution
freezes at sufficiently low temperatures describing the process of liquid
to glass transition. The time evolution of the probability distribution
towards the equilibrium distribution during subsequent aging describes
the primary relaxation of a glass.

\bigskip
\bigskip

{\bf 1. Introduction}\newline
The difficulty in defining the glassy state~\cite{1} at the
phenomenological (or macroscopic) level lies not only in the
description of the spatial arrangement of constituent atoms which can
qualitatively be described by terms such as non-crystalline or
amorphous or absence of long range order~\cite{9}. An important further
challenge is in describing the aging behavior of glasses over long
periods of time during which a glass gradually changes from being
solid-like at short times to liquid-like at long times. This
time-dependent behavior has generated a lively debate as to how to
describe a glass: Is it a solid that creeps over long times, or a
frozen liquid that relaxes with time, or something in between like a
visco-elastic solid? Questions such as
{\textquotedblleft}Do cathedral glass windows flow?{\textquotedblright}
are still blowing in the wind~\cite{2,3}. 

\bigskip

In this paper, we present a semi-microscopic physics-based view of glass
forming systems. We refer to this view as the potential energy landscape
(PEL) view. This PEL-view is based on the firmly established and widely
accepted statistical thermodynamic framework of glass forming systems
that provides an understanding of the liquid and supercooled liquid
states, of the liquid to glass transition, and of the glassy 
state ~\cite{9,4,5}.  It is evident that the PEL-view is
not a hypothesis or a model, since it is as much a reality as are
the interatomic interactions in a system. The PEL-view contains the
essential physics that rationalizes the generic features of structural
glass-forming systems.  Important universal features of these systems are~\cite{9,6}:

\begin{itemize}
\item 
Super-Arrhenius increase of the viscosity of the
liquid state with decrease in temperature,
\item 
Non-linear, nonexponential, non-Arrhenius and
spatially heterogeneous primary (or $\alpha$) relaxation in the
supercooled liquid state,
\item 
Liquid to glass transition temperature (sometimes referred to as the fictive 
temperature) decreasing with
increasing observation time during cooling of a
liquid,
\item 
Absence of latent heat in liquid to glass
transition,
\item 
Rounded discontinuous drops in heat capacity and in compressibility 
during the liquid to glass transition, 
\item 
Prigogine-Defay ratio value greater than unity during the liquid to glass transition,
\item 
Second order properties in the glass state (such as the heat capacity) similar to
those in the corresponding crystalline state,
\item 
History dependent properties in the glassy state,
\item 
Aging of glassy state properties at temperatures below the glass transition temperature.
\item 
Near-Arrhenius secondary (also called Johari-Goldstein or $\beta$)
relaxation.
\end{itemize}

\bigskip

The PEL-view was introduced by Goldstein in his classic 1969
paper~\cite{7}.  The underlying concepts and theories based on the
PEL-view have been much reviewed although largely in the condensed
matter physics literature~\cite{9,4,8,10}.  For this reason, details are
not presented in this paper. The purpose of this paper is therefore to
present just a simplified introduction to the concepts and aspects of PEL
that are of particular interest to structural glasses, to describe how
the PEL-view rationalizes important features of glass forming systems,
and to formulate a new description of real (sometimes termed kinetic or
non-equilibrium) glasses in terms of basis glass states that are uniquely
determined from the PEL.

\bigskip

The organization of the paper is as follows. In section 2 basic concepts
and thermodynamics of a potential energy landscape are described. Section
3 describes the dynamics of a system in the PEL since it is the dynamics that
define different states (e.g., the liquid and glassy states) of a system
and transitions between them such as the liquid to glass transition. In
section 4, we discuss the equilibrium (i.e., ergodic) liquid state in
terms of PEL features. Section 5 discusses the liquid to glass transition
and the non-equilibrium (broken-ergodic) glassy state. In section 6,
we develop the concept of basis glass states and provide a description
of a real glass as a superposition of basis glass states. In section
7, we discuss time dependent processes (e.g., structural relaxation)
that are always present in a real glass-forming system. This paper is
concluded in section 8.\\[5mm]

{\bf 2. The Potential Energy Landscape
\newline
}
The PEL is a highly rugged $3N+1$ dimensional continuous
hyper-surface representing the total potential energy ($\Phi$) 
of a system containing a large number ($N$) of atoms
(sometimes referred to as particles) as a function of atomic
configurations specified by $3N$ atomic coordinates:

\begin{equation}
\Phi =\Phi({\textbf r}_{1},{\textbf r}_{2},{\dots},{\textbf r}_{N})
\label{eq1}
\end{equation}

\noindent
where ${\textbf r}_i$ is the coordinate of particle $i$. (Note that for
non-thermal systems like hard spheres, $\Phi$ is a constant and therefore
it is more appropriate to consider the free energy landscape~\cite{11}. In
this paper, we only consider thermal systems.) The PEL is fully determined
from the knowledge of the atomic interactions, composition and volume of
the system. We emphasize that the PEL is independent of temperature ($T$). Temperature
enters the picture only via the equilibrium probability
to find the system at a certain point in the configuration space, i.e.,
the Boltzmann weight of a configuration:

\begin{equation}
{\rm Prob}({\textbf r}_{1},{\textbf r}_{2},\dots,{\textbf r}_N) \propto
\exp \left[-\frac{\Phi({\textbf r}_{1},{\textbf r}_{2},\dots,{\textbf r}_N) 
}{kT}\right] \quad ,
\label{eq2}
\end{equation}
\bigskip

\noindent
where $k$ is the Boltzmann constant.

Since the PEL contains the full information about the
potential energy of the system, the thermodynamic properties of the
latter can in principle be obtained in an exact manner. \ However, the
interesting aspect of the PEL approach is related to the fact that many
properties of the glass-forming systems can be obtained by a relatively
small set of geometric properties of the PEL, thus reducing tremendously
the information necessary to describe the system~\cite{4,12}.  Important
geometric features of a PEL are~\cite{5,10,13,14}:

\begin{itemize}
\item 
The inherent structures (IS): The local energy
minima, called ISs, represent stable configurations of the system and
provide important information on the atomic level structure (e.g., the
radial distribution function). Note that the set of ISs also contains
the ground state of the system which in most cases corresponds to a
crystalline phase, although exceptions to this are known~\cite{15}. In the
following we will assume that the system is a good glass-former and
avoids all crystalline configurations, i.e., we consider only the
non-crystalline configurations in the PEL.

\item 
The basins of attraction (BA): The set of states
that upon energy minimization via steepest descent lead to a given IS
constitute the basin of attraction (BA) of that IS. The PEL can be
partitioned in a unique way into distinct BA{\textquoteright}s. \ At
low temperatures and short times, the system is confined in these BAs
and hence they are important for rationalizing the vibrational
properties of the system.

\item 
The curvatures of the PEL at the ISs: Knowledge of these curvatures
provides direct information on the vibrational density of states of
the system.

\item 
The number of ISs as a function of energy: This dependence determines
if the glass-former is strong (weak energy dependence i.e., small
configurational heat capacity) or fragile (strong energy dependence i.e.,
large configurational heat capacity).

\item 
The saddles or transition states~\cite{16,17}: These are stationary
states of a PEL having at least one negative principal curvature. Their
respective energies are related to the non-vibrational
(i.e., configurational or structural) relaxation in the system at low
temperatures. 

\item 
The steepest descent paths connecting an IS to a neighboring IS via
the common transition state: Such a path corresponds to the reaction
coordinate for structural relaxation at low temperatures.

\item 
The distribution of the barrier energy of a PEL~\cite{18} (defined as
energy of the transition state minus the energy of the starting IS):
These barriers play an important role in governing the dynamics of
transitions among different ISs. This dynamics is thermally activated
around temperatures near and below the glass transition temperature.

\end{itemize}

These features of a PEL refer to the properties of \textit{individual}
ISs and BAs. However, it is also of great interest to go beyond
this local view and in particular to see how the different BA's are
connected to each other on larger scales. Previous studies~\cite{10}
have given evidence that the PEL is not just a random landscape
in which BA{\textquoteright}s and IS{\textquoteright}s are patched
together in a random manner, but that instead there is a hierarchy in
which IS{\textquoteright}s with similar energy are clustered together
giving rise to {\textquotedblleft}meta-basins{\textquotedblright} (MB),
i.e., a collection of IS{\textquoteright}s that are relatively easily
accessible to each other but surrounded by larger barriers from other
IS's. As we will see below, these MBs are important for understanding
the behavior of the glassy state.

The MBs give rise to another partition of the PEL, different from the BA
partition, that we will refer to as the ``meta-basin partition''. It is
a unique partition of a PEL and results when inter-IS transitions of the
system are sequentially turned off starting from the highest barrier and
going down to some threshold barrier height. When a sufficient number
of inter-IS transitions are turned off in this manner, a PEL uniquely
partitions into a large number of non-overlapping components (namely MBs)
among which no transitions exist involving barrier heights greater than
the threshold barrier height. Lower barrier transitions 
paths remain active only within MBs. It should be noted that in general each MB contains
many ISs and there are exponentially extensive number of MBs. This
MB-partition and the IS-structure of MBs (which IS belongs to which MB)
are uniquely determined for a PEL and play a critical role in defining
the nature of glass transition and of the glassy state when the PEL
dynamics is thermally activated~\cite{10}.

We also mention that it is possible to extend these concepts
to isothermal-isobaric systems using the enthalpy landscape
formalism~\cite{18A,18B}.  However, because the primary goal of this paper
is to introduce the new concept of basis glass states (see section~6)
in a clear and rigorous manner, we limit here the discussion to 
systems in the canonical ensemble. This in no way limits the general validity of
the results and conclusions of this paper.

An important point to emphasize is that while it may be extremely
challenging (and may not even be possible) to calculate in an exact
manner any or all these geometric features for a given PEL, in principle
all these features are uniquely and completely determined for a PEL. In
other words, these features are properties of a PEL and therefore of the
system. From the theoretical point of view this is an important point
since it allows development of relatively simple (abstract) models that
describe at least some of the features of a glass-forming system. The
well-known ``random energy model'' is an example of such an abstract
PEL~\cite{19}.\\[5mm]

{\bf 3. Dynamics in a PEL\newline }
The configuration point of a system in the canonical ensemble (i.e.,
in contact with a thermal bath at temperature $T$) moves around
rapidly on the PEL. The above mentioned properties of the PEL,
distribution of the IS-energy, barrier heights, the MB-partition,
etc., have a direct consequence on how the system explores the
PEL. Previous studies~\cite{4,10,20} have shown the existence
of three dynamical regimes: Above the onset-temperature $T_0$,
the typical values of the IS-energies that the system explores are
independent of temperature and the relaxation dynamics of the system is essentially
exponential with time and has an Arrhenius temperature dependence. For
temperatures in the range $T_0 > T > T_c$, where $T_c$ is the critical
temperature of the mode-coupling theory~\cite{21}, the average IS-energy
decreases~\cite{22} but the system still mainly probes the PEL close to
the saddle points~\cite{16,23,24}. The relaxation becomes increasingly
non-exponential in time and its temperature dependence super-Arrhenius
as the temperature is reduced. In practice this means that a typical
configuration of the particles has a dynamical matrix (i.e. the
matrix describing the local curvatures of the PEL) which has at least
one negative eigenvalue. For mean-field like systems this is an exact
result~\cite{25} but simulations have shown that realistic glass-forming
systems show the same behavior. Below $T_c$, the system mainly resides
deep inside the BAs and explores the local minima in the PEL making only
rare transitions between BAs of neighboring local minima. Hence at these
temperatures the dynamical matrix has only positive eigenvalues. Thus,
only for temperatures below $T_c$ one expects the system dynamics to
be thermally activated and, at sufficiently low temperatures, transport
coefficients such as the diffusion coefficient and viscosity gradually
revert back towards Arrhenius behavior~\cite{9}. With decreasing
temperature, the system spends more time in lower energy ISs and the
barrier heights of the transitions increase leading to rapid slowing
down of the dynamics.

This temperature dependence of the dynamics of the system in the PEL
finds its correspondence in the relaxation dynamics of the glass-former
when it is described by the usual dynamic observables, such as the
mean-squared displacement of a particle or the intermediate scattering
function~\cite{26}. For supercooled liquids that show a glassy dynamics,
i.e., for temperatures below $T_0$, it is well known that the dynamics can
be decomposed into two parts~\cite{9}: A fast one which corresponds to
the rapid vibrational motion of the particles inside the cage formed by
their neighbors and a much slower motion that is related to the escape
of the particles from this cage and that corresponds to the structural
relaxation of the system. While the fast dynamics occurs on the time
scale of atomic vibrations, i.e., picoseconds, the relaxation slows
down rapidly with decreasing $T$ and hence will be orders of magnitudes
slower. As a consequence, we have in the supercooled regime a strong
separation of these two time scales. This fact can be used to formulate
an approximate description of the dynamics of the system in the PEL: The
fast vibrational motion of the particles corresponds to a dynamics inside the
BA of an IS. Transitions between neighboring ISs in the same MB (i.e.,
intra-MB transitions) can be associated with the Johari-Goldstein or
$\beta$-relaxation, i.e.  the broad peak seen in dielectric measurements
that, at low temperatures, is at frequencies higher than the $\alpha$-peak
and which shows an Arrhenius dependence on $T$~\cite{26b}. Transitions between ISs in
different MB (i.e., inter-MB transitions) reflect the $\alpha$-relaxation
of the system~\cite{27}. 

The stochastic dynamics for the inter-IS dynamics can be approximated
by a master equation for the probability $p_i(t)$ for the system to be
present at time $t$ in the $i$-th IS-basin~\cite{28}:

\begin{equation}
\frac{dp_i}{dt}=\sum_j [w_{ij}p_j(t) - w_{ji}p_i(t)] \quad .
\label{eq3}
\end{equation}

\noindent
Here, $w_{ij}$ represents the $T$-dependent transition rate from the
$j$-th IS to the $i$-th IS. Note that by using this master equation one
neglects memory effects, i.e., one assumes that the separation of time
scales between vibrations and relaxations is large enough so that the
system loses its memory of previous IS transitions.

Measurements of an observable property of a system are carried out
over some finite period of time, which in the following we will denote
{\textquotedblleft}observation time{\textquotedblright} $(t_0)$.
Whether or not a measured property represents the equilibrium value
of the system depends on the relative magnitudes of two different time
scales: the time of observation which is an external time scale imposed
by the observer, and an internal time scale $(\tau)$ characterizing
the kinetics of relaxation processes taking place in the system. Since
vibrational times are extremely short, on the order of picoseconds,
vibrational properties are usually in equilibrium for typical observation
times of an experiment. When the typical structural relaxation time
is significantly smaller than the observation time, the system is in
thermodynamic equilibrium and is called ergodic.

A non-ergodic system is termed broken-ergodic when the
distribution of relaxation times exhibits slow and fast modes such that

\begin{equation}
\tau_{\rm fast} < t_0 < \tau_{\rm slow}
\label{eq4}
\end{equation}

\noindent
In such a situation, a system is in equilibrium
(during the observation time) with respect to the fast modes but is
frozen, \ i.e., non-ergodic, with respect to the slow modes. This
implies that the configuration point of the system is confined (during
the observation time) with some probability in a subspace (such as a
MB) of the configuration space and is not able to explore a significant
portion of the PEL. Note that this probability over the subspace will
depend on the history of the system. Palmer~\cite{29} was the first to
formulate the statistical mechanics of such broken-ergodic systems.

For glass forming systems, it is useful to consider also the
configurational changes in terms of MB-dynamics, i.e., the time dependence
of metabasin probabilities $P_\alpha(t)$ for the system configuration
point to be present in the $\alpha^{\rm th}$-MB at time $t$. These
MB-probabilities are defined as follows:

\begin{equation}
P_\alpha(T,t) =\sum_{i \in \alpha} p_{i}(T,t) \quad .
\label{eq5}
\end{equation}

\noindent
Here the sum is over all ISs that constitute the $\alpha^{\rm th}$-MB.
Note that since the equilibrium value of $p_i$ depends on temperature,
$P_\alpha$ will depend on $T$ as well. The MB-dynamics can be
approximately described by a suitably aggregated master equation for
MB-probabilities~\cite{30}.

From the formal point of view, in the above equations, one has considered
the PEL of only {\it one} system. One might thus wonder how in a real experiment
on a macroscopic sample the probabilities $p_i$ and $P_\alpha$ come into
play since at any time only one IS and MB will be occupied. In practice a macroscopic
sample can be divided in many macroscopically small (but microscopically
large) pieces, each of which will have the same properties as the original
system. Since the individual pieces, or sub-systems, are independent from each other,
apart from small corrections due to presence of common interfaces,
the probability that a given sub-sample is in the $i$-th IS-basin is
given by $p_i$. So, a property of the entire sample is given by the
weighted sum of the property of the sub-samples. Thus, it is indeed the
probability distribution $\{p_i\}$ that determines the behavior of the
macroscopic sample.\\[5mm]

{\bf 4. The liquid state\newline}
For a specified observation time $t_0$, a system at sufficiently high
temperatures is able to explore the entire configuration space and the
MB-probabilities take their equilibrium values $P_\alpha({\rm eq},T)$.  This
equilibrium (ergodic) state corresponds to the liquid or the supercooled
liquid state (from now on, we simply refer to it as the liquid state).

A large configurational entropy is a defining characteristic of the
liquid state. The configurational entropy, $S_c({\rm Liq},T)$, of the
liquid state is expressed by the Gibbs entropy equation~\cite{31}:

\begin{equation}
S_c({\rm Liq},T)=-k\sum_i p_i({\rm eq},T)\ln p_i({\rm eq},T) \quad .
\label{eq6}
\end{equation}

\noindent
Here k is the Boltzmann's constant and the summation is over all ISs
of the entire PEL. Note that the probabilities $p_i({\rm eq},T)$ depend
on temperature and hence at low $T$ only the IS with low energy will be
populated. As a consequence, $S_c({\rm Liq},T)$ decreases when temperature
is lowered~\cite{32}. According to Eqn.~(\ref{eq6}), the configurational entropy
will be positive for {\it all} temperatures and will vanish only at
absolute zero (assuming that the degeneracy of the ground states is
sub-exponentially extensive).

The configurational entropy of a liquid can also be
expressed in terms of the configurational entropy of MBs:

\begin{equation}
S_{c}({\rm Liq},T)=\sum_\alpha P_\alpha ({\rm eq},T) S_{c,\alpha}
({\rm eq},T)+I({\rm eq},T) \quad .
\label{eq7}
\end{equation}

\noindent
Here, $S_{c,\alpha}$ is the configurational entropy of the $\alpha
$-th MB and is defined in a manner similar to eqn (\ref{eq6}) but with
probabilities normalized for the $\alpha $-th MB~\cite{33,34}:

\begin{equation}
S_{c,\alpha}({\rm eq},T)=
-k\sum _{i\in \alpha }\left(\frac{p_i(T)}{P_\alpha(T)}\right)
\ln \left(\frac{p_i(T)}{P_\alpha(T)}\right)
\label{eq8}
\end{equation}

The quantity $I({\rm eq},T)$, called complexity, is defined as
follows~\cite{9,35}:

\begin{equation}
I({\rm eq},T)=-k \sum_\alpha P_\alpha ({\rm eq},T)
\ln P_\alpha ({\rm eq},T) \quad .
\label{eq9}
\end{equation}

\noindent
In Eqn.~(\ref{eq9}) the sum is over the all MBs. Complexity represents
a global property of the liquid and is that part of the total
configurational entropy which is associated with the exploration of
different MBs of the PEL. On the other hand, a property of a MB such
as $S_{c,\alpha}$ represents a local property (that of a single MB
of the PEL). Similar to the configurational entropy, the magnitude of
the complexity decreases as $T$ is reduced. It is therefore tempting
to argue that at low temperatures the configurational entropy and the
complexity are the same, implying that at these $T$'s the system is
confined in the BA of just {\it one} IS and thus the MB and the BA are
identical~\cite{32}. However, recent studies indicate that this is not
the case, i.e. that even at very low temperatures the configurational
entropy is larger than the complexity, or in other words, the MB have
a non-trivial structure even at very low $T$~\cite{36}. It is possible
that for some systems, the complexity may vanish at some low but finite
temperature (similar to but not necessarily equal to the Kauzmann
temperature~\cite{37,38}) below which a system is trapped in the ground
state MB.

Viscous flow in liquids is a manifestation of $\alpha$-relaxation (i.e.,
inter-MB transitions) and the viscosity is proportional to the average
$\alpha$-relaxation time. A non-Arrhenius behavior of viscosity is a
characteristic feature of glass forming liquids~\cite{39} and it signals
the increase of the barrier heights with decreasing temperature. There
is considerable debate about its low temperature behavior: does the
viscosity diverge at some finite temperature (in the manner of the
Vogel-Fulcher (VF) eqn.~\cite{40}), does it revert to the Arrhenius
behavior with divergence only at absolute zero~\cite{41}, or does
it follow a different functional form with no divergence at finite
temperature~\cite{42}. One reason for this debate is the absence of
viscosity data at low temperatures near where the divergence is expected
based on the VF equation. Empirical equations (such as the MYEGA
eqn.~\cite{43}, B\"assler eqn.~\cite{44}, or the ECG eqn.~\cite{45})
with the same (or even lower~\cite{44}) number of fitting parameters as the
VF equation and that do not exhibit any divergence have been found to fit
the experimental data just as well as the VF equation. For the viscosity
to diverge at a finite temperature, the barrier must also become infinite
at the divergence temperature.  Since this is not the case for the PEL
of systems that have a finite interaction range (which implies an upper
bound for the barrier height), it is impossible to support a Vogel-Fulcher
type of temperature dependence at low temperatures. Note that a maximum
(but finite) barrier height at low temperatures will imply a change to
an Arrhenius behavior at low temperatures as has been reported by some
investigators~\cite{41}.

In the PEL formalism, the increase in barrier heights with decreasing
temperatures is largely due to the fact that at low $T$ the liquid
starts to explore ISs with low energies. These IS's are connected to
neighboring IS with barriers that become increasingly large since the
local arrangement of the particles becomes more and more optimized and
breaking up this local arrangement becomes increasingly difficult. A
consequence of this picture is that the temperature dependence
of the barrier height should be approximately proportional to the
configurational heat capacity of the liquid. This correlation between
temperature dependence of the barrier height and the configurational
heat capacity of the liquid is indeed supported by experimental data
and has been noted by several investigators~\cite{46,47}. For example,
pure silica (a strong glass forming liquid) exhibits nearly Arrhenius
behavior and has extremely small configurational heat capacity while OTP
(a fragile liquid) is highly non-Arrhenius and has a large configurational
heat capacity~\cite{48}.\\[5mm]

{\bf 5. The glass transition and the glassy state}\newline
Upon cooling a liquid, the structural relaxation dynamics slow down
and inter-IS transitions begin to freeze sequentially starting with
the highest barrier. At some low temperature, when a sufficient
number of inter-IS transitions are blocked, the PEL partitions into
MBs that are no longer mutually accessible on the time scale of the
experiment. This MB-partitioning of the PEL corresponds thus to the
liquid to glass transition. The partitioned state is broken-ergodic
and corresponds to the glass state~\cite{49,50}. The temperature at which
the partitioning occurs depends on the cooling rate and is called the
fictive temperature $T_f$ of the glass. Technically, $T_f$ should be called the
glass transition temperature. Unfortunately, the term glass transition
temperature customarily refers to the temperature where the viscosity
is $10^{12}$~ Pa.s (i.e., average relaxation time about 100 sec) and
hence $T_f$ is also named ``kinetic glass transition temperature''. We
emphasize that in the liquid to glass transition the system is trapped
in MBs and not in individual basins of attraction. Thus, even in the
glass state the system will undergo some localized intra-MB relaxation dynamics
at low temperatures, the details of which we will discuss below.

At the glass transition, $P_\alpha$, i.e, the probability that the system
is in the $\alpha$-MB, freezes at a value equal to that with which the
liquid was exploring the MB immediately prior to the glass transition:

\begin{equation}
P_\alpha ({\rm glass}\,| T_f)=P_\alpha ({\rm eq},T_f) \quad .
\label{eq10}
\end{equation}

We emphasize that while the values of the MB-probabilities $P_\alpha$
depend on the time of observation via the fictive temperature, the
geometry of the MB-partition does not.

It is easy to see that the energy of the system does not change during
glass transition since the basin probabilities $p_i$ are the same in
glass and in the liquid at $T_f$ . However, this is not true for the
entropy. The entropy of a glass with fictive temperature $T_f$ and at a
bath temperature $T$ is the thermal average of the entropy of the MBs
weighted at $T_f$~\cite{33}:

\begin{equation}
S({\rm glass},T,T_f)= \sum_\alpha P_\alpha ({\rm eq},T_f)S_\alpha
({\rm eq},T) \quad .
\label{eq11}
\end{equation}

\noindent
Note that on the right hand side, $S_\alpha ({\rm eq},T)$ is evaluated
at $T$ and not at $T_f$, because each MB remains ergodic, so only the
weights $P_\alpha$ depend on $T_f$.

The comparison of liquid and glass entropy expressions shows that
there is a net entropy loss during the discontinuous liquid to glass
transition. (Note that the discontinuous nature of glass transition is
an intrinsic feature of partitioning of the PEL and is not associated
with cooling rate or observation time used.) This is simply because the
average configuration space volume of a MB is much smaller than that of
the configurational space volume of the entire system. Eqs.~(\ref{eq7})
and (\ref{eq11}) show that the magnitude of the entropy loss is given
by the complexity evaluated at $T_f$:

\begin{equation}
S({\rm Liq},T_f)-S({\rm glass},T=T_f,T_f)=I(T_f) \quad .
\label{eq12}
\end{equation}

\noindent
This loss of entropy entails a drop in the heat capacity during the
liquid to glass transition. Further, since there is no change in the
energy at the glass transition, the free energy, $F$, of the system
increases discontinuously at $T_f$ with respect to the free energy of
the liquid state:

\begin{equation}
F({\rm glass},T=T_f,T_f)-F({\rm liq},T_f)=T_f I(T_f) \quad .
\label{eq13}
\end{equation}

It must be emphasized that a glass transition is not an equilibirum
thermodynamic transition but a dynamic transition. Because of its unusual
features (no change in energy, a loss in entropy, and an increase in the
free energy), the glass transition is sometimes also referred to as a
``Random First Order Transition''(RFOT)~\cite{51,52}.

With no change in energy (i.e., no latent heat), the notion of a loss
in entropy may appear to violate the second law.  However, in reality
there is no violation because a liquid to glass transition is not a
spontaneous process. Instead, it is a process caused by an externally
imposed constraint of finite observation time.  Unlike this view,
the conventional picture assumes entropy continuity during a glass
transition~\cite{53,54}. This difference in the two views has a direct
consequence on the magnitude of the residual entropy, i.e., the entropy
at absolute zero. According to the PEL-view, since no inter-IS transitions
are kinetically possible at $T = 0$, the configurational entropy vanishes
at absolute zero whereas in the traditional view, glasses do have a
finite residual entropy~\cite{55}.  So far, it has not been possible
to test experimentally the entropy change during a glass transition
because entropy is not an experimentally observable quantity (unlike
energy, it is not a dynamic variable) and there is no defined way of
calculating the entropy change from the experimentally measurable heat
capacity in a non-equilibrium, non-spontaneous process like the liquid
to glass transition.\\[5mm]

{\bf 6. Basis glass states and real glasses\newline}
Since the IS-structure of a MB is fixed and there is no overlap among MBs,
one can use the MBs to define a ``basis-glass state''.  In other words,
for a basis-glass based on the ${\alpha}-$MB, we have $P_\alpha({\rm
Basis}) = 1$ and the weights for all other MBs are zero. Note that
we distinguish between the terms MB and basis-glass state: A MB is
a region in configuration space defined by the MB-partition while a
basis-glass represents a (hypothetical) state of the system with some
probability distribution for the ISs within the MB. By definition,
a basis-glass cannot exhibit ${\alpha}$-relaxation but it can undergo
$\beta$-relaxation. It is worth emphasizing that a basis-glass may exhibit
some temperature dependence since not all ISs within a MB have the same
energy. This in turn makes the position and the width of the $\beta $-peak
depend on temperature. The configurational entropy of a basis-glass
is the same as the configurational entropy of the corresponding MB and
hence is much less than the total configurational entropy of the liquid.
Because of its low configurational entropy, a basis-glass will behave
like a solid.

Let us consider a real glass which we symbolically denote by $G({\rm
real},T, t| T_f)$, formed at a fictive temperature $T_f$ and aged at
temperature $T$ for some time $t$. For times longer than the
observation time used during prior cooling, the MBs are no longer
dynamically isolated and hence the system starts to make some inter-MB
transitions even at temperatures lower than $T_f$. This causes the
MB-probabilities $\{P_\alpha (T_f)\}$ to depend on time $t$ as well as
the temperature $T$ of the sample. Let us denote the probability that the
system is found in the the MB $\alpha$ by $P_\alpha(T,t|T_f)$.  The real
glass can thus be considered as a linear combination of basis-glasses
$\{G_\alpha \}$ with $T$ and time dependent weights $P_\alpha(T,t|T_f)$:

\begin{equation}
G({\rm real},T,t| T_f)=\sum_\alpha
P_\alpha (T,t|T_f)G_\alpha ({\rm basis},T) \quad ,
\label{eq14}
\end{equation}

\noindent
where $G_\alpha ({\rm basis},T)$ is a glass that is confined in MB $\alpha$.

Thus, in the PEL view, a real glass is described by the set of (time and
temperature dependent) MB-probabilities,$\{P_\alpha(T,t|T_f)\}$.  As a
real glass ages, it evolves irreversibly through a continuum of different
glassy states. This implies, for example, that an aged glass and a
freshly formed glass of the same system are not the same but two different
glasses. It should be noted that the time dependence of a real glass is
contained primarily in the time-dependence of the MB-probabilities. On
the other hand, the basis-glasses are the primary reason for the solid
like behavior of real glasses at short observation times.

Any observable property of a real glass is thus given by the
average of its values over all basis-glass states. For example, the
average energy, $\langle E \rangle$, of a real glass is the
average energy of basis-glasses:

\begin{equation}
\langle E \rangle ({\rm real},T,t| T_f)=\sum_\alpha
P_\alpha (T,t|T_f)\langle E \rangle_\alpha ({\rm basis},{\rm eq},T) 
\label{eq15}
\end{equation}

It should be noted that because a real glass is a superposition of
basis-glass states, a real glass is truly heterogeneous given by a
mosaic of basis-glass states in the physical 3-dimensional space. The size
of the mosaic ``tiles'' is controlled by the small amorphous-amorphous
interfacial energy between different basis-glass states~\cite{56}.\\[5mm]

{\bf 7. Structural relaxation in the glass state\newline}
Real glasses exhibit two types of relaxation processes: fast secondary
($\beta $) due to intra-MB transitions and slow primary ($\alpha$) due
to inter-MB transitions. For short times, the inter-MB transitions remain
frozen and the system exhibits only secondary relaxation without altering
the MB-probabilities since the barriers associated with inter-MB transitions
are larger than the barriers for inter-IS transitions within the same
MB. This is consistent with the observation that the range of barrier
heights for the secondary relaxation is on average less than that of
the $\alpha$-relaxation~\cite{57}.  Note that the identity of a real glass
does not change during secondary relaxation. Over longer times, inter-MB
transitions allow the system to relax slowly, i.e. the glass will age.

The PEL-view readily explains several observed features of the
$\alpha$-relaxation in glasses. For example, one should note that because
a real glass is a mosaic of basis-glasses, the $\alpha$-relaxation
process is spatially heterogeneous. During $\alpha$-relaxation,
the MB-probabilities $P_\alpha(t)$ evolve with time from their
initial value $P_\alpha(\rm eq,T_f)$ in the glassy state towards
their final equilibrium value $P_\alpha(\rm eq,T)\}$ in the liquid
state. This time dependence of the $\alpha$-relaxation is nonexponential
because of the distribution of inter-MB barrier heights that become
active during relaxation. While the non-exponentiality may resemble a
stretched-exponential behavior~\cite{58}, in general the $\alpha$-relaxation
may not be a true stretched exponential because there is an upper bound
to $\alpha$-barrier heights which is mathematically not the case in a
stretched exponential function.  

We emphasize that real glasses exhibit $\alpha$-relaxation even for
temperatures much less than $T_f$ provided the observation time is
sufficiently increased as is the case in aging experiments. The
initial stages of $\alpha$-relaxation in a glass (generally referred
to as ``iso-structural'' relaxation) have been investigated in a
variety of glasses~\cite{59}. The iso-structural flow is nearly Arrhenius
and its activation energy is observed to be significantly less than
the activation energy for the equilibrium $\alpha$-relaxation in the
corresponding liquid at $T_f$, the observed values of the ratio are in
the range from 0.13 - 0.68~\cite{59}. This can be understood by realizing that
the inter-MB barriers which begin to unfreeze during the early stages of
iso-structural relaxation in a glass are precisely the ones that were
the last (and hence the smallest) to freeze at $T_f$ just before the
MB-partitioning during the prior cooling of the liquid.

According to the MB-based view advanced here, the conceptual foundations
of the physics underlying relaxation processes are simple and follow the
principle of ``last to freeze is first to relax''. The roadmap essentially
consists of sequential unfreezing of thermally activated transitions
starting with the smallest barriers. In principle, one could develop
the mathematical formalism of relaxation dynamics rigorously by solving
the non-isothermal non-linear time evolution of the 
associated master equation for the MB-probabilities
starting from the moment when the liquid was last in equilibrium. This
will require a detailed knowledge of the barrier height distribution
of the hyper-dimensional PEL and is beyond the purpose and scope of this
paper.\\[5mm]

{\bf 8. Concluding Remarks\newline}
In this paper, we have argued that the PEL view of glassy forming
systems provides a rational framework to describe the observed universal
behavior of kinetic glass transition and of structural relaxation.
This framework is rooted in the fundamental constructs of
statistical mechanics and an assumption of thermally activated stochastic
dynamics in the PEL. In addition to rationalizing the observed behavior,
this formalism leads to new concepts such as basis-glass states (that
are unique to a system) and provides a description of real glasses
as a superposition of basis-glass states. Further, it leads to new
consequences such as the loss of complexity (a portion of the liquid
configurational entropy) during the glass transition and of zero residual
entropy in the glassy state. Last but not least the concept of MBs and
basis glass states allows one to rationalize the fact that even deep in
the glass state the particles are not completely frozen but still can
undergo a local dynamics that is not of vibrational type and which is
related to the Johari-Goldstein $\beta$-relaxation.

\bigskip

\textbf{Acknowledgments:}

\bigskip

Prabhat Gupta is extremely grateful to all his
colleagues, collaborators, coauthors, and friends in the glass
community for their generous support and encouragement throughout his
career. Walter Kob acknowledges the support of ANR grant
ANR-15-CE30-0003-02.

\end{document}